\documentclass[10pt]{iopart}

\usepackage{iopams}  
\usepackage{graphicx}
\usepackage{url}
\usepackage{booktabs}
\usepackage{multirow}
\usepackage{dcolumn}
\usepackage{xspace}
\newcolumntype{d}[1]{D{.}{.}{#1}}

\def\ket#1{|\,#1 \,\rangle}

\def\ms{\,\mathrm{ms}}

\def\tref#1{table~\ref{#1}}

\def\uext{$u_\mathrm{ext}$}

\def\fresult{$353\,638\,794\,073\,800.332(91)\,$Hz\xspace}

\begin{document}

\title[Absolute frequency measurement of a $^{176}$Lu$^+\,(^{3}\mathrm{D}_1)$ standard]{Absolute frequency measurement of a $^{176}$Lu$^+\,(^{3}\mathrm{D}_1)$ standard against the NRC-FCs2 fountain with $2.6\times10^{-16}$ uncertainty}


\author{$^{1,2}$K. J. Arnold, $^4$Bin Jian, $^1$Zhao Zhang, $^1$Qi Zhao, $^1$Qin Qichen, $^1$N. Jayjong, $^1$M. D. K. Lee, $^4$Scott Beattie, $^{1,3}$M. D. Barrett}
\address{$^1$Centre for Quantum Technologies, National University of Singapore, 117543, Singapore}
\address{$^2$Temasek Laboratories, National University of Singapore, 117411, Singapore}
\address{$^3$Department of Physics, National University of Singapore, 117551, Singapore}
\address{$^4$Metrology Research Centre, National Research Council Canada, Ottawa, ON K1A 0R6, Canada}
\ead{cqtkja@nus.edu.sg; bin.jian@nrc-cnrc.gc.ca}
\vspace{10pt}

\begin{abstract}
We report an improved absolute frequency measurement of a $^{176}$Lu$^+\,(^{3}\mathrm{D}_1)$ optical frequency standard, operated at the National University of Singapore (NUS), evaluated via a remote link to the NRC-FCs2 caesium fountain primary frequency standard. Operating a single ion clock with 94.2\% uptime over 10 days, and using an ambiguity-resolved precise point positioning (PPP-AR) link over the Global Positioning System (GPS), we determine an absolute frequency of \fresult at a fractional uncertainty of $2.6 \times 10^{-16}$. This agrees with our previous result, which underpins the CIPM recommended frequency value, and reduces the uncertainty by a factor of 3.6.
\end{abstract}

%
\vspace{2pc}

\submitto{\MET}

%
\maketitle
%
\ioptwocol

\section{Introduction}
The International System of Units (SI) currently defines the second through the ground-state hyperfine transition of $^{133}${Cs}. This SI second is best realized by caesium fountain primary frequency standards (PFS), with inaccuracies near $1\times10^{-16}$~\cite{weyers2018advances,beattie2020first}. Optical atomic clocks have surpassed microwave standards in both instability and accuracy, with many systems now reporting systematic uncertainties in the $10^{-18}$ regime and below~\cite{mcgrew2018atomic,takamoto2020test,sanner2019optical,aeppli2024clock,marshall2025high,arnold2025optical,hausser2025in+,lindvall202588}. The Consultative Committee for Time and Frequency (CCTF) has therefore established a roadmap toward redefining the SI second using an optical transition~\cite{dimarcq2024roadmap}. A mandatory milestone is the realization of multiple, independent, high-accuracy absolute frequency measurements of optical clocks, currently recognized as secondary representations of the second (SRS), against independent caesium fountain PFS~\cite{weyers2018advances,beattie2020first}, together with demonstrated agreement among those measurements.

The $^1\rm S_0 \leftrightarrow {^3\rm D_1}$ transition in $^{176}${Lu}$^+$ is a competitive frequency reference, having recently demonstrated a record-low systematic uncertainty of $1\times10^{-19}$ and agreement between two independent systems at a statistically limited uncertainty of $5\times10^{-19}$~\cite{arnold2025optical}. The first absolute frequency measurement of this standard yielded $353\,638\,794\,073\,800.35(33)$~Hz, a fractional uncertainty of $9.2 \times 10^{-16}$~\cite{zhang2025absolute}. While this established an important baseline and underpins the Comit\'e international des poids et mesures (CIPM) 2025 recommended frequency~\cite{cipm_2025}, it was limited by satellite-link instability and optical-clock uptime.

Here we report a high-accuracy absolute frequency measurement of a single-ion $^{176}${Lu}$^+$ ($^3\rm D_1$) standard in Singapore, evaluated against the NRC-FCs2 fountain in Canada~\cite{beattie2020first} through an ambiguity-resolved precise point positioning (PPP-AR)~\cite{petit2022continuous,Jian_2023} remote link. Operating the standard with 94.2\% uptime over the 10-day campaign suppresses the dead-time extrapolation uncertainty, leaving a total uncertainty primarily limited by the NRC-FCs2 fountain~\cite{beattie2020first}. The resulting $2.6\times10^{-16}$ is a 3.6-fold reduction relative to the previous measurement~\cite{zhang2025absolute} and the first measurement of $^{176}${Lu}$^+$ to fall below the roadmap criterion of $3\times10^{-16}$ for continuity with the Cs-based SI second~\cite{dimarcq2024roadmap}. The result, \fresult, is in excellent agreement with the current CIPM recommended frequency, further establishing $^{176}${Lu}$^+$ as a reliable frequency reference.

\section{NUS-Lu$^+$ System and Systematics \label{sec:ExpSys}}

The $^{176}\mathrm{Lu}^{+}$ optical frequency standard operated at the National University of Singapore (NUS), together with the local measurement system, is described in detail in~\cite{zhang2025absolute}. Here we highlight only the key differences: the upgraded ion trap and ultra-stable reference cavity.

Experiments are performed with a single $^{176}\mathrm{Lu}^{+}$ ion in a linear Paul trap denoted Lu-1~\cite{arnold2025optical}. Compared to earlier work~\cite{zhiqiang2023176lu+}, this trap was cleaned and rebuilt to address an anomalously high heating rate, and the rf resonator was replaced to lower the rf drive frequency and increase confinement. The trap was operated at an rf drive frequency of 9.4~MHz, with measured secular trap frequencies of $[0.188, 1.15, 1.23]$~MHz. 

The reference cavity has been upgraded to a 30-cm ultra-low-expansion (ULE) optical cavity (Stable Laser Systems) with a finesse of $190\,000$ at 1550~nm. A 1550~nm fiber laser (NKT) is frequency-offset locked to the cavity using a wideband electro-optic modulator (EOM). A frequency comb transfers the stability of the 1550~nm laser to the 848~nm clock laser: the comb repetition rate is stabilized by phase-locking the beat note between the comb and the 1550~nm laser to a direct digital synthesizer (DDS), and the 848~nm laser is then phase-locked to the comb at a fixed frequency offset. The DDS that sets the offset of the 1550~nm laser from the cavity continuously compensates the linear ULE cavity drift of approximately $48\,\mathrm{mHz}\,\mathrm{s}^{-1}$, and applies additional phase-continuous corrections for steering during $\mathrm{Lu}^+$ clock servo operation.

The $\mathrm{Lu}^{+}\,(^{3}{\rm D}_1)$ standard frequency is defined as the average of the three optical transitions from $\ket{g}\equiv\ket{^1{\rm S}_0,7,0}$ to the hyperfine states $\ket{F}\equiv\ket{^3{\rm D}_1,F,0}$ with $F=6,7,$ and $8$. This average is conveniently realized using a hybrid microwave and optical interrogation~\cite{kaewuam2020hyperfine}. Hyper-Ramsey spectroscopy~\cite{yudin2010hyper,hyperRamsey2012peik} is additionally incorporated to suppress the probe-laser-induced ac Stark shift. The timing parameters used here for the hyper-Ramsey (HR) hyperfine-average (HA) interrogation sequence~\cite{zhiqiang2023176lu+,arnold2025optical} are $\tau_L = 6$~ms for the optical $\pi$-time, $\tau = 3$~ms for both microwave $\pi$-times, and $T_R = 50$~ms for the total Ramsey time. For these parameters the Ramsey fringe has nearly full contrast, and the $\mathrm{Lu}^{+}$ clock operated with a quantum-projection-noise-limited fractional frequency instability of $4.0\times10^{-15}/\sqrt{\tau/\mathrm{s}}$.

A hydrogen maser (HM, Microchip MHM-2010) directly references all synthesizers and the GNSS receiver, and is linked to the $\mathrm{Lu}^+$ 848~nm optical clock laser frequency via the frequency-comb measurement scheme described in~\cite{zhang2025absolute}. The maser frequency offset was zeroed before the measurement campaign, and no additional steering was applied during the campaign.

The $\mathrm{Lu}^+$ clock systematic uncertainty budget for this campaign is summarized in \tref{lusys}. The data was collected with the `Lu-1' apparatus shortly before the two-clock comparison and full systematics evaluation reported in~\cite{arnold2025optical}, but under markedly different operating conditions. That work used correlation spectroscopy between the two optical references, which supported Ramsey interrogation times of 5~s and longer, far beyond the laser coherence time. The present single-clock measurement instead uses a 50~ms Ramsey time limited by laser coherence---a hundredfold reduction---and this much shorter interrogation time is the main reason the two budgets differ, as it renders the probe-induced shifts comparatively more significant. Nevertheless, the total systematic uncertainty remains dominated by the gravitational redshift, which is assessed as reported in~\cite{zhang2025absolute}.

The microwave ac-Zeeman shift arises from imbalanced $\sigma^\pm$ polarization components of the microwave fields which off-resonantly couple to $\Delta m = \pm 1$
Zeeman transitions. The polarizations of both microwave
fields were characterized, but not optimized, at the time of the campaign by measuring the relative coupling strength of the $\Delta m = [-1,0,1]$ microwave transitions and the shift evaluated as in~\cite{zhiqiang2023176lu+,arnold2025optical}. It is noted that after this campaign, the microwave horns were mounted on rotation mounts which allow for easy balancing of $\sigma^\pm$ components to suppress this shift in the future. 

Other systematic shifts are as evaluated in~\cite{arnold2025optical}. The second-order Doppler shift from thermal motion is substantially reduced, since ion heating is negligible over the 50~ms Ramsey time.
\begin{table}[t]
\centering
\footnotesize
\caption{\label{lusys} Uncertainty budget of the Lu$^+$ clock for this campaign. All values are in fractional frequency units of $10^{-18}$ Hz/Hz. }
\begin{tabular}{l d{2} d{2}}
\toprule
Effect & \multicolumn{1}{c}{Shift} & \multicolumn{1}{c}{Uncertainty} \\
\midrule
Quadratic Zeeman shift & -151.29 & 0.10 \\
ac-Zeeman (microwave)  & 88.5    & 3.4  \\
ac-Zeeman (rf)         & 0.27    & 0.01 \\
Blackbody radiation    & -1.37   & 0.10 \\
Second order Doppler   & -0.07   & 0.04 \\
Microwave coupling     & 0.0     & 0.13 \\
Residual quadrupole    & 0.03    & \llap{$<$}0.01 \\
\midrule
Total (ex redshift)    & -63.9   & 3.4  \\
Gravitational redshift & 1717    & 36   \\
\midrule
Total (with redshift)  & 1653    & 36   \\
\bottomrule
\end{tabular}
\end{table}

\section{NRC-FCs2 System and Systematics  \label{sec:FCs2}}
NRC-FCs2 is a caesium fountain primary frequency standard at the National Research Council Canada (NRC).  Since the first accuracy evaluation in 2020~\cite{beattie2020first}, it has been reporting to the Bureau international des poids et mesures (BIPM) for regular calibrations of International Atomic Time (TAI).  In 2024, several of the dominant systematic shifts were re-evaluated, resulting in a reduction of the total uncertainty to $1.1\times10^{-16}$~\cite{beattie2025advancements}.  The evaluation of four systematic effects of NRC-FCs2 were updated in this campaign, i.e., the cold collisions, the blackbody radiation (BBR), the second-order Doppler, and the $m=1$ component of the distributed cavity phase (DCP) shifts.  

NRC-FCs2 operated by alternating the atomic density between high and low to extract the collision-free frequency in real-time.  The evaluation of the associated cold collisional frequency uncertainty requires the knowledge of the atomic density ratio and its uncertainty as well as the frequency difference between the two densities~\cite{szymaniec2007cancellation, szymaniec2010primary}.  In 2023, the atom cloud density has been characterized using absorption imaging and the density ratio uncertainty was reduced by 10-fold~\cite{beattie2023characterization}.  During this campaign, the fountain operated near the collisional shift cancellation point~\cite{szymaniec2007cancellation}.  This resulted in a cold collisional shift uncertainty of $2.4\times10^{-17}$ for this campaign.      

A lower uncertainty scalar polarizability~\cite{rosenbusch2007blackbody} and a re-evaluated lower temperature uncertainty of the fountain drift tube were used to calculate the BBR shift and the associated uncertainty compared to that in~\cite{beattie2025advancements}, leading to a reduction of the BBR uncertainty from $0.61\times10^{-16}$ to $0.36\times10^{-16}$.  The DCP $m=1$ component was evaluated with an uncertainty of $0.4\times10^{-16}$ using absorption imaging through the vertical axis of the NRC-FCs2 physics package in 2024~\cite{beattie2023characterization}.  This uncertainty was enlarged by a factor of two for this campaign (February 2025)  due to the aging of the vertical laser beam alignment.  The second order Doppler effect of NRC-FCs2 was also included here which contributes a frequency shift of less than $1\times10^{-17}$ (absolute value) and a negligible uncertainty~\cite{ashby2021falling}.   

During the frequency measurement of NRC-FCs2, the C-field value and the drift tube temperature were logged in real-time to evaluate the second order Zeeman and the BBR shifts, respectively.  All the other systematic shifts and uncertainties of NRC-FCs2 are based on the 2024 re-evaluation~\cite{beattie2025advancements}.  The overall systematic uncertainty of NRC-FCs2, $u_{\mathrm {B[FCs2]}}$, is $1.17\times10^{-16}$ during this campaign and table~\ref{fcs2sys} presents the systematic uncertainty budget.  

The microwaves at 9.192 GHz for the operation of NRC-FCs2 were provided using a commercial synthesizer (Spectra Dynamics Cs-1) referenced to an active hydrogen maser at NRC (VM1, BIPM code 1400307).  The fountain instability was limited by the local oscillator with an Allan deviation of $\sigma_y(\tau)=1.65\times10^{-13}/\sqrt{\tau/s}$.

\begin{table}[t]
\centering
\footnotesize
\caption{\label{fcs2sys} Uncertainty budget of NRC-FCs2 in this campaign. All values are in fractional frequency units of $10^{-16}$ Hz/Hz. }
\begin{tabular}{l d{4} d{4}}
\toprule
Effect & \multicolumn{1}{c}{Shift} & \multicolumn{1}{c}{Uncertainty} \\ \midrule
Zeeman shift                  & 724.25  & 0.2    \\
Blackbody radiation           & -163.12 & 0.36   \\
Second order Doppler          & -0.0982 & 0.0004 \\
Cold collisions               & 0       & 0.24   \\
DCP $m=0$                     & 0.06    & 0.34   \\
DCP $m=1$                     & 0       & 0.8    \\
DCP $m=2$                     & 0.04    & 0.02   \\
Microwave lensing             & 0.6     & 0.2    \\
Microwave leakage             & 0       & 0.2    \\
Synchronous phase transients  & 0       & 0.5    \\
Background gas collisions     & 0       & 0.08   \\
Microwave spectral impurities & 0       & 0.1    \\
Rabi and Ramsey pulling       & 0       & 0.1    \\
Cavity pulling                & 0       & 0.1    \\
Majorana transitions          & 0       & 0.1    \\
AC Stark (light)              & 0       & 0.001  \\
\midrule
Total (ex redshift)           & 561.73  & 1.17   \\
Gravitational redshift        & 104.55  & 0.03   \\
\midrule
Total (with redshift)         & 666.28  & 1.17   \\
\bottomrule
\end{tabular}
\end{table}

\section{Frequency Link \label{sec:freqlink}}

The absolute frequency of the NUS Lu$^+$ ion clock transition was determined against the SI second realized by the NRC-FCs2 primary frequency standard via a GPS PPP-AR link during an interval between the modified Julian date (MJD) 60717.5 and 60727.5 (10 d).  The GPS link connects the flywheel oscillators, i.e. the hydrogen masers, at the two labs with no interruptions.  In this section, we show the details of the frequency measurement chain from the Lu$^+$ clock to the fountain.

\begin{figure}
\centering\includegraphics[trim=0cm 0cm  0cm 0cm, clip=true, keepaspectratio=true,  angle=0, width=8.0 cm ]{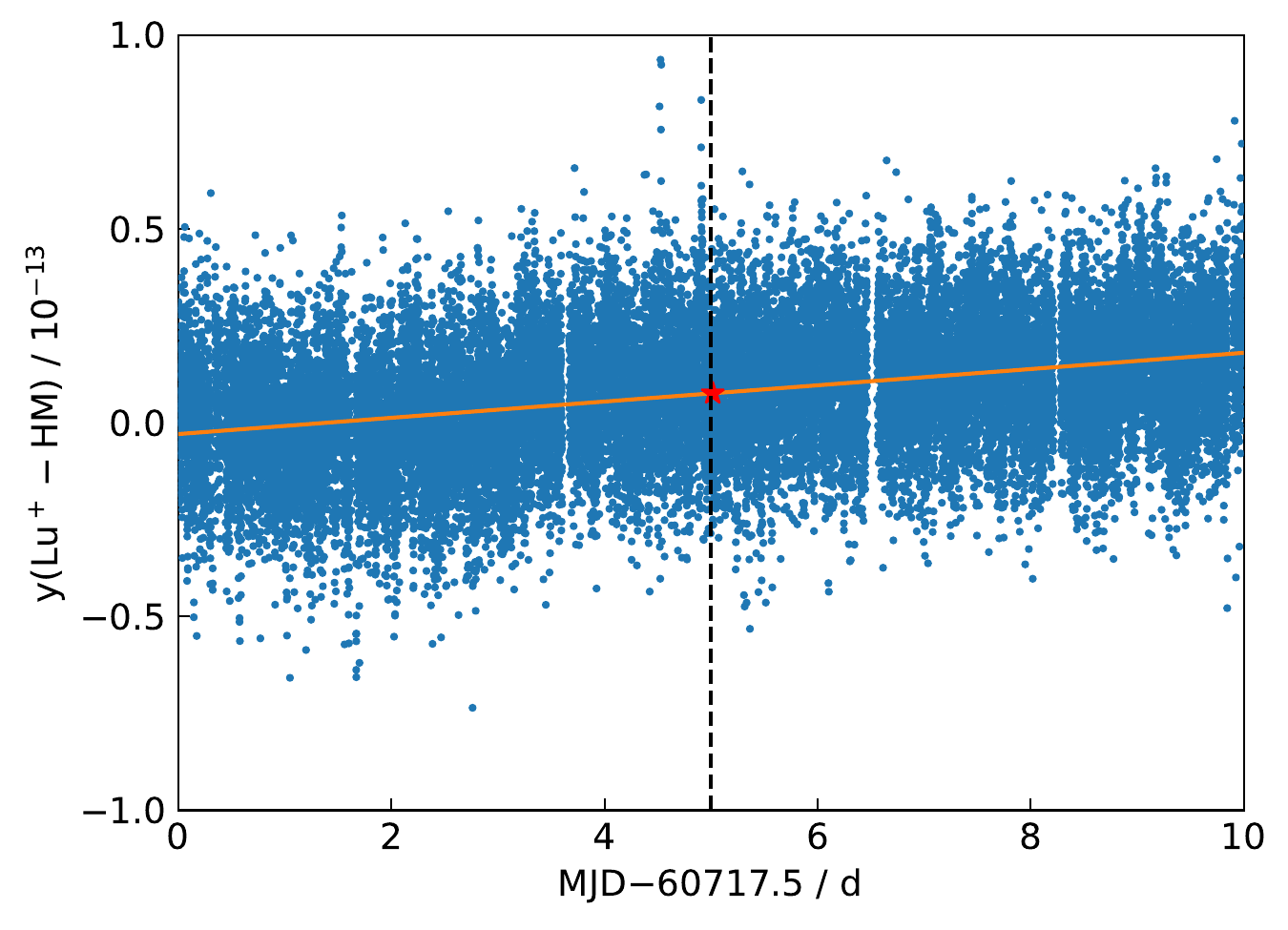}%
\caption{\label{fig:yLu+_HM} The fractional frequency offset between the Lu$^+$ ion clock and the NUS HM.  The systematic shifts have been corrected.  The blue points are the frequency measurement data binned for the 30-s GPS time grids.  The orange line is the weighted linear fit to the data to estimate the HM drift.  The red star shows the weighted arithmetic mean of the data.  The vertical dashed line denotes the center of this campaign on MJD 60722.5.      
}
\end{figure}

\subsection{Lu$^+$ ion - HM and extrapolation}
Figure~\ref{fig:yLu+_HM} shows the frequency difference between the Lu$^+$ and the NUS HM during this campaign.  The 2025 CIPM recommended frequency for the $^{176}$Lu$^+$ clock transition was used and all the systematic shifts have been corrected~\cite{cipm_2025}.  The Lu$^+$ clock achieved an uptime of 94.2\% during the measurement campaign. 

The data in figure~\ref{fig:yLu+_HM} shows the frequency drift of the NUS HM during the 10-d period.  A linear fit was used to estimate the HM drift rate as $-2.10(3)\times10^{-15}$/d.  The average frequency offset between the Lu$^+$ ion clock and the HM was first calculated as the weighted mean of the data in figure~\ref{fig:yLu+_HM} as $7.558\times10^{-15}$ with the barycenter on MJD 60722.5174.  The duration of the frequency measurement in each bin was used as the weight.  This frequency offset was then corrected to the center of the campaign on MJD 60722.5 using the HM drift leading to the value of $\overline{y}(\mathrm{Lu^+-HM})$ of $7.521\times10^{-15}$.  The statistical uncertainty from the Lu$^+$ ion clock, $u_\mathrm {A[Lu^+]}$, is calculated as $4\times10^{-18}$ based on the Lu$^+$ ion clock instability of $4.0\times10^{-15}/\sqrt{\tau/s}$ and its total measurement time.

\begin{figure}
\centering\includegraphics[trim=0cm 0cm  0cm 0cm, clip=true, keepaspectratio=true,  angle=0, width=8.0 cm ]{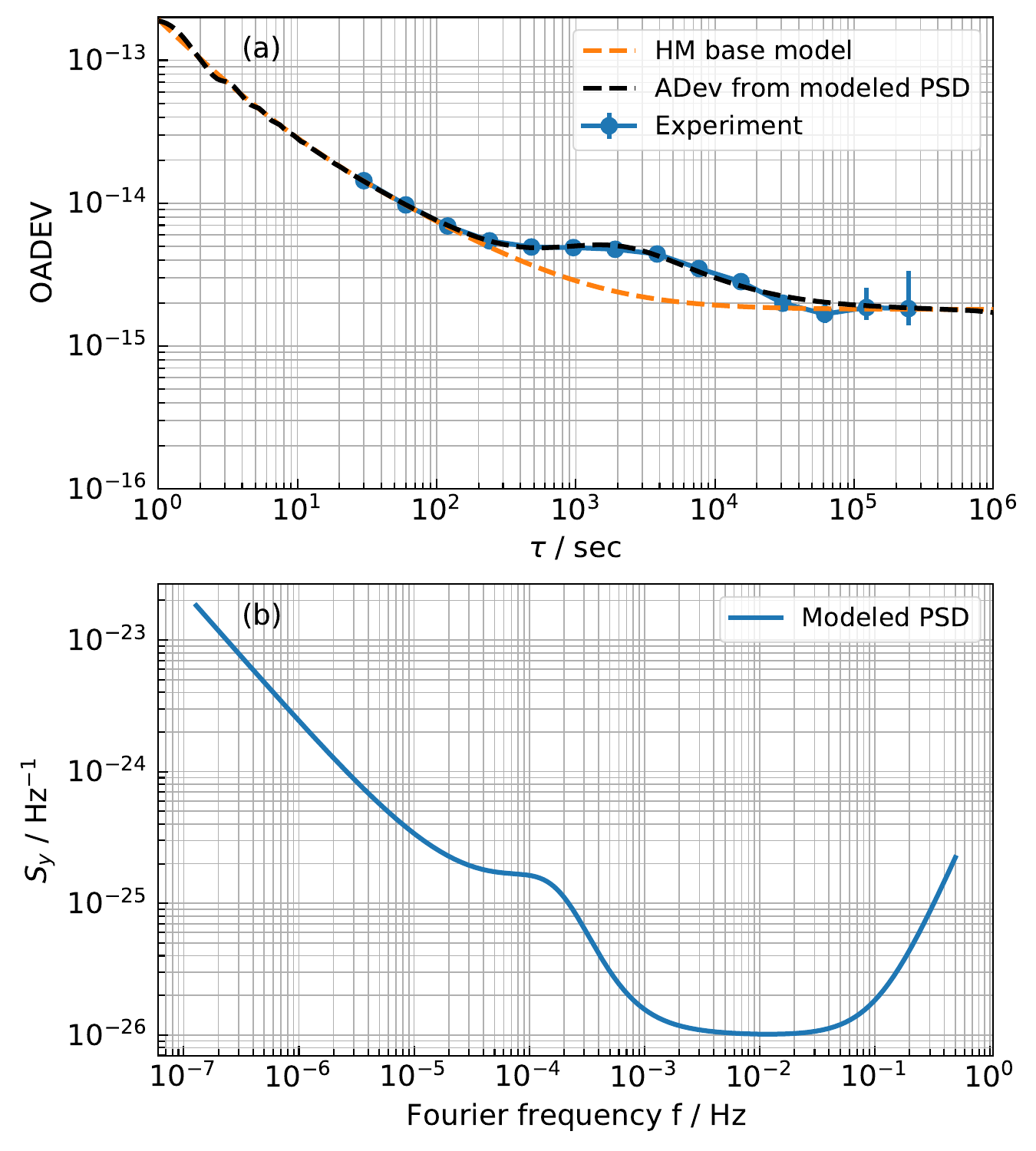}%
\caption{\label{fig:hm_allan} Noise model of the NUS HM as a flywheel oscillator for the Lu$^+$ ion clock.  (a) Allan deviation of the NUS HM showing the de-drifted frequency data (blue circles), base maser model (dashed orange line), and a modified noise model (dashed black line).  (b) The single-sided PSD for the HM including a Lorentzian contribution accounting for the plateau in the Allan deviation.      
}
\end{figure}

The average frequency offset between the Lu$^+$ ion clock and HM is subject to the effect of the ion clock measurement downtime during which the maser frequency noise was sampled.  The related extrapolation uncertainty, \uext, is estimated using the Fourier transform method~\cite{grebing2016realization, lindvall2025coordinated, dawkins2007considerations}, which uses the modeled power spectral density (PSD) of the maser and the Fourier transform of a weighting function dependent on the uptime distribution of the ion clock over the 10-d interval.  Figure~\ref{fig:hm_allan} (a) shows the Allan deviation of the NUS HM calculated using the data shown in figure~\ref{fig:yLu+_HM}, de-drifted.  A base model of the Allan deviation (dashed orange line) comprised power-law contributions from white phase modulation (WPM), white frequency modulation (WFM), and flicker frequency modulation (FFM), with values of $1.8\times10^{-13}/\tau$, $7\times10^{-14}/\sqrt{\tau}$, and $1.8\times10^{-15}$, respectively. The corresponding PSD model was calculated from its Allan deviation as $h_2f^2+h_0+h_{-1}/f$, where $h_2 = 8.5\times10^{-25}\,$Hz$^{-3}$, $h_0 = 9.8\times10^{-27}\,$Hz$^{-1}$, and $h_{-1} = 2.3\times10^{-30}$ correspond to the noise contributions due to WPM, WFM, and FFM, respectively.  This model failed to capture the behavior of the maser between times 300 and 20\,000 s, where a bump at the level of $5\times10^{-15}$ is observed in the Allan deviation. In order to model this behaviour, the PSD was modified by adding a Lorentzian function $A/(1+(f-f_0)^2/\delta f^2)$ with $A = 1.3\times10^{-25}\,$Hz$^{-1}$, $f_0 = 1.0\times10^{-4}\,$Hz, and $\delta f = 1.5\times10^{-4}\,$Hz.  Figure~\ref{fig:hm_allan} (b) shows the combined PSD of the base maser noise model and the Allan deviation bump.  The resulting Allan deviation converted from the combined PSD maser noise model is shown in figure~\ref{fig:hm_allan} (a) as the dashed black line and shows good agreement with the experimental data~\cite{riley2008handbook}.  The maser extrapolation uncertainty, $u_{\mathrm {ext}}$, was then calculated using the Fourier transform method with the combined maser PSD model and the weighting function of the ion clock uptime distribution as $7.4\times10^{-17}$.  The uncertainty due to the drift of the maser, $u_\mathrm d$, is calculated from the maser drift uncertainty ($3.3\times10^{-17}$/d) and the offset between the evaluation interval center and the data barycenter.  The value of $u_\mathrm d$ is $1\times10^{-18}$.          

\subsection{PPP-AR}

A GPS PPP-AR link between the NUS receiver and a NRC receiver (NRCC) was used to link the flywheel oscillator hydrogen masers in the two labs.  Both receivers are Septentrio PolaRx5TR.  The PPP-AR link was formed by concatenating the differences of the daily solutions of the two receivers generated from the Natural Resources Canada (NRCan) Canadian Spatial Reference System (CSRS) online service using the NRCan final products~\cite{nrcan_csrs}.  The possible day boundary phase discontinuities (DBPD) in the PPP-AR link were fixed using the reference method as described in Ref~\cite{jian2024continuous}.  The reference PPP-AR link was generated with the same two receivers by processing the PPP-AR solutions as a batch with the NRCan rapid products which maintain the day boundary phase continuity.  

\begin{figure}
\centering\includegraphics[trim=0cm 0cm  0cm 0cm, clip=true, keepaspectratio=true,  angle=0, width=8.0 cm ]{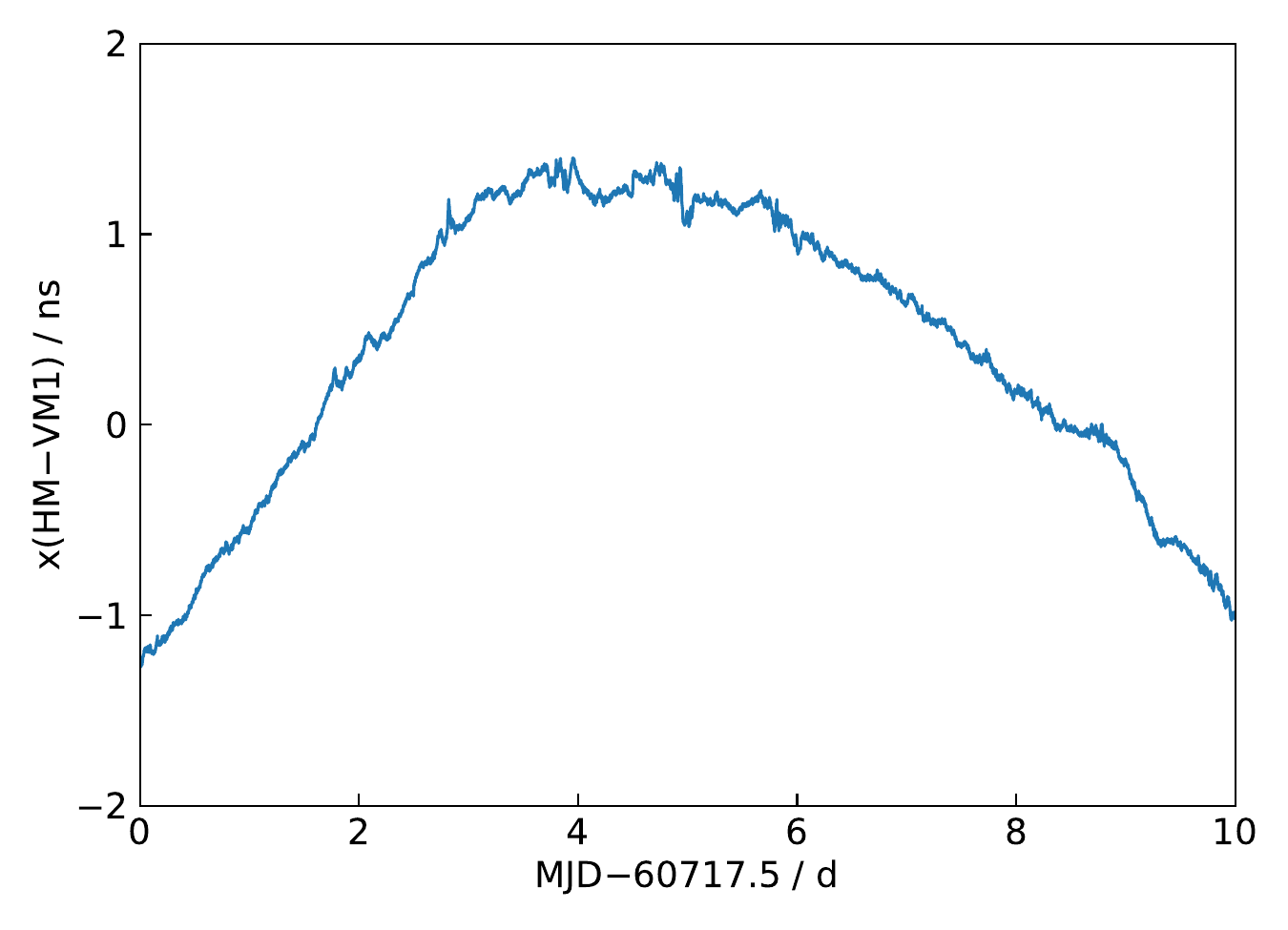}%
\caption{\label{fig:link} Phase difference of flywheel oscillator masers between the two labs.  A constant linear fit (frequency offset) has been removed.  }
\end{figure}

The PPP-AR link gives the phase difference between the reference clocks of the receivers in the two labs, i.e. HM for the NUS lab and UTC(NRC) for NRC.  The phase data between the NRC flywheel oscillator, VM1, and UTC(NRC) was measured using a TimeTech 100 MHz phase comparator (PCO) and then sampled at the 30-s GPS timestamp grids to combine with the PPP-AR link.  The phase data of the hydrogen masers between the two labs linked through PPP-AR is shown in figure~\ref{fig:link}.  The average frequency offset between the two masers was calculated as $\overline{y}(\mathrm {HM-VM1})=(x_{\mathrm {end}} - x_{\mathrm {start})}/T = 1.0048\times10^{-13}$.  Here $x_{\mathrm{start/end}}$ is the phase difference between the NUS HM and the NRC VM1 at the start and the end of the evaluation interval and $T$ is the measurement duration of 10 d.  The frequency transfer uncertainty (FTU) of PPP-AR is estimated as $1\times10^{-16}$ using the formula $1\times10^{-15}/T$ with $T$ in days~\cite{petit2022continuous, jian2024continuous}.

The rf distribution uncertainty at the NRC lab was estimated using the Allan deviation of the 100 MHz PCO data between VM1 and UTC(NRC).  A noise floor at the level of $6\times10^{-17}$ is reached after several hours of averaging, which is used as the rf distribution uncertainty.  This uncertainty is omitted for the NUS lab since the receiver was directly referenced to the HM signals.        
  
\subsection{NRC-FCs2 - VM1}

\begin{figure}
\centering\includegraphics[trim=0cm 0cm  0cm 0cm, clip=true, keepaspectratio=true,  angle=0, width=8.0 cm ]{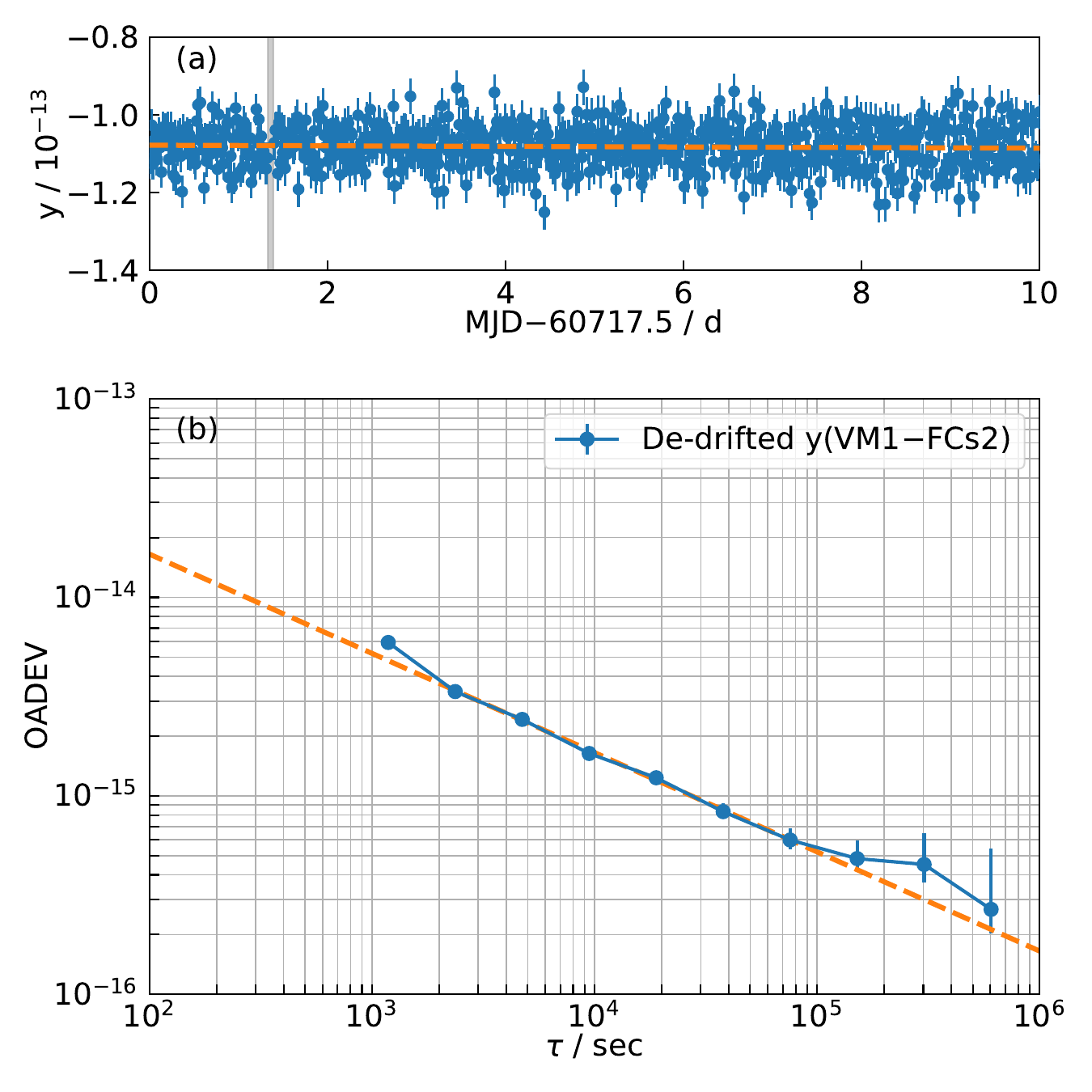}%
\caption{\label{fig:fcs2} Frequency measurement of VM1 against NRC-FCs2.  (a) Frequency offset between VM1 and NRC-FCs2, $y(\mathrm{VM1-FCs2})$.  The systematic shifts of NRC-FCs2 have been corrected.  The dashed line is a weighted linear fit to the data.  The vertical gray line denotes the measurement downtime of 1.5 h due to scheduled maintenance.  (b) Allan deviation of NRC-FCs2 as calculated from the data shown in (a), de-drifted.  The first data point starts with $\tau_0$ of 1181 s as the duration of the high and low atomic densities for the fountain to extract the collisional cancellation frequency.  The dashed line shows the modeled fountain instability of $1.65\times10^{-13}/\sqrt{\tau}$. }
\end{figure}

NRC-FCs2 operated normally during the frequency measurement campaign.  A scheduled maintenance interruption of $\sim$1.5 hours happened on MJD 60718.86.  Figure~\ref{fig:fcs2} (a) shows the frequency measurement of VM1 against NRC-FCs2 after all the systematics have been corrected for the fountain as described in section~\ref{sec:FCs2}.  Figure~\ref{fig:fcs2} (b) shows the Allan deviation of the fountain measurement after removing the VM1 frequency drift of $7.6\times10^{-17}$/d.  We treat the fountain data as a continuous measurement throughout the 10-d interval.  This is justified by the high uptime operation of NRC-FCs2 (99.4\%) and the low drift of the NRC VM1.  The average frequency offset of VM1 with respect to NRC-FCs2, $\overline{y}(\mathrm{VM1-FCs2})$, is then calculated as the arithmetic mean of $-1.0805\times10^{-13}$ and the statistical uncertainty $u_{\mathrm {A[FCs2]}}$ is estimated as $1.8\times10^{-16}$ using the fountain noise model as shown in figure~\ref{fig:fcs2} (b) and the total measurement time. 

\section{Results and Discussion}

\begin{table}[t]
\centering
\footnotesize

\caption{\label{tab:results} Frequency links from NUS Lu$^+$ ion clock to NRC-FCs2. All values are in fractional frequency units of $10^{-18}$ Hz/Hz.  All fractional offsets are referenced to the CIPM 2025 recommended frequency of the $^{176}$Lu$^+$ clock transition, $353\,638\,794\,073\,800.35\,$Hz~\cite{cipm_2025}.  NRC-FCs2 realizes the SI second with all known biases of the fountain and of the frequency link corrected, so that $\overline{y}(\mathrm{Lu^+-FCs2})$ is the fractional deviation of the frequency measured here from that recommended value. }

\lineup
\begin{tabular}{lrlrc}
\toprule

 & $\overline{y}$~~ & Uncert. &  & uptime \\ \midrule
\multirow{4}{*}{Lu$^+-$HM} & \multirow{4}{*}{7\,521} & $u_\mathrm {B [Lu^+]}$ & 36 & \multirow{4}{*}{94.2\%} \\
 & & $u_\mathrm {A [Lu^+]}$ & 4 & \\
 & & $u_{\mathrm {ext}}$ & 74 & \\
 & & $u_{\mathrm d}$ & 1 & \\ \midrule
 \multirow{2}{*}{HM$-$VM1} & \multirow{2}{*}{$100\,481$} & $u_{\mathrm {lnk}}$ & 100 & \multirow{2}{*}{100\%} \\
 & & $u_{\mathrm {rf}}$ & 60 &  \\ \midrule
 \multirow{2}{*}{VM1$-$FCs2} & \multirow{2}{*}{$-108\,054$} & $u_\mathrm{B [FCs2]}$ & 117 & \multirow{2}{*}{99.4\%} \\
 & & $u_\mathrm{A [FCs2]}$ & 178 & \\ \midrule
 & & Total uncert. & & \\ \midrule
 \multirow{2}{*}{Lu$^+-$FCs2} & \multirow{2}{*}{-52} & $u_{\mathrm B}$ & 122 &  \\
 & & $u_{\mathrm A}$ & 226 & \\ 
\bottomrule
\end{tabular}
\end{table}

Table~\ref{tab:results} summarizes the frequency offset and the uncertainties of each measurement link and presents the $^{176}$Lu$^+$ ion clock transition frequency measured against NRC-FCs2 through PPP-AR: $\overline{y}(\mathrm {Lu^+-FCs2})=-0.52\times10^{-16}$, with total statistical and systematic uncertainties of $2.3\times10^{-16}$ and $1.2\times10^{-16}$ in fractional frequency, respectively.  The corresponding absolute frequency of the $^{176}$Lu$^+$ ion clock transition is thus \fresult.  This result agrees well with the previous absolute frequency measurement of the NUS Lu$^+$ ion clock via a GPS PPP link with a reduction of the overall uncertainty by a factor of 3.6 reaching to the low level of $10^{-16}$ and limited by the NRC-FCs2 primary frequency standard~\cite{zhang2025absolute}.  

The NRC-FCs2 instability of $1.65\times10^{-13}/\sqrt{\tau/s}$ contributes the dominant uncertainty of $1.8\times10^{-16}$ in this campaign.  The instability could readily be reduced by more than a factor of two if optically synthesized microwaves were used as demonstrated in~\cite{marceau2025absolute}.  In addition, this uncertainty can also be reduced by extending the measurement campaign from the current 10-d duration to a few tens of days.  This would also significantly improve the PPP-AR link FTU as it scales inversely with the measurement duration.  If we assume that the optically synthesized microwaves were used and the measurement campaign lasted 30 days (all other uncertainties unchanged), a total uncertainty of $1.7\times10^{-16}$ can be achieved which is lower than the systematic uncertainty of many Cs primary frequency standards for TAI calibrations.   

The sub-$10^{-16}$ uncertainty of the extrapolation of the NUS HM, $u_{\mathrm {ext}}$, is attributed to the high uptime ratio of the Lu$^+$ ion clock (94.2\%) and the relatively even distribution of the measurement downtime.  With the current NUS HM (see figure~\ref{fig:hm_allan} for its noise model), it is important to measure as continuously as possible to minimize the extrapolation uncertainty not only because of the non-power law noise, but also because of the relatively high FFM noise at the level of $1.8\times10^{-15}$.       

Finally, as one may note from figure~\ref{fig:yLu+_HM}, the NUS HM frequency drift changed around MJD 60720.5.  It is interesting to repeat the data analysis as outlined in section~\ref{sec:freqlink} for two periods before and after MJD 60720.5.  The frequency difference value is a weighted average of the results of the two analysis intervals with the weight calculated using the statistical uncertainty for each interval.  The final frequency difference of Lu$^+$ versus NRC-FCs2 determined in this way is $\overline{y}(\mathrm {Lu^+-FCs2})=-0.8\times10^{-16}$, with total statistical and systematic uncertainties of $2.4\times10^{-16}$ and $1.2\times10^{-16}$, respectively.  This result agrees with that analyzed using the 10-d interval with similar overall uncertainty.  We quote the result from the single 10-d interval analysis as the result for this work.

\section{Conclusions \label{sec:conclusion}}

To summarize, we present an absolute frequency measurement of the NUS Lu$^+$ ion clock against the NRC-FCs2 Cs primary frequency standard.  The result is \fresult, which agrees with the previous measurement. The overall uncertainty of 0.091 Hz demonstrates a 3.6-fold improvement in accuracy from the previous measurement.  In the future, by operating the fountain with optically stabilized microwaves, it is feasible to reduce the overall uncertainty to $1.7\times10^{-16}$ for a measurement of 30 days.    

\ack
We thank Brian Donahue, Elyes Hassen, and Simon Banville for providing the SPARK software and access to the NRCan PPP-AR products which were used to generate the reference PPP-AR link. 

This project is supported by the National Research Foundation, Singapore through the National Quantum Office, hosted in A*STAR, under its National Quantum Engineering Programme 3.0 Funding Initiative (W25Q3D0007) and under its Centre for Quantum Technologies Funding Initiative (S24Q2d0009).\\

\section*{Data availability}

Data sets that support the findings of this study are available upon reasonable request from the corresponding authors.

\section*{References}
\bibliographystyle{iopart-num}

\providecommand{\newblock}{}

\end{document}